\documentclass[aps,prd,showpacs,amssymb,twocolumn,nofootinbib]{revtex4}
\usepackage{graphicx}
\begin{document}
\title{Microcausality and Quantum Cylindrical Gravitational Waves}

\author{J. Fernando \surname{Barbero G.}}
\email[]{jfbarbero@imaff.cfmac.csic.es} \affiliation{Instituto de
Matem\'aticas y F\'{\i}sica Fundamental, Centro de F\'{\i}sica Miguel A.
Catal\'{a}n, C.S.I.C., Serrano 113bis-121, 28006 Madrid, Spain}
\author{Guillermo A. \surname{Mena Marug\'an}}
\email[]{mena@imaff.cfmac.csic.es} \affiliation{Instituto de
Matem\'aticas y F\'{\i}sica Fundamental, Centro de F\'{\i}sica Miguel A.
Catal\'{a}n, C.S.I.C., Serrano 113bis-121, 28006 Madrid, Spain}
\author{Eduardo J. \surname{S. Villase\~nor}}
\email[]{eduardo@imaff.cfmac.csic.es} \affiliation{Instituto de
Matem\'aticas y F\'{\i}sica Fundamental, Centro de F\'{\i}sica Miguel A.
Catal\'{a}n, C.S.I.C., Serrano 113bis-121, 28006 Madrid, Spain}

\date{February 25, 2003}

\begin{abstract}
We study several issues related to the different choices of time
available for the classical and quantum treatment of linearly
polarized cylindrical gravitational waves. We pay especial
attention to the time evolution of creation and annihilation
operators and the definition of Fock spaces for the different
choices of time involved. We discuss also the issue of
microcausality and the use of field commutators to extract
information about the causal properties of quantum spacetime.
\end{abstract}


\maketitle


\section{\label{Intro}Introduction}

The quantization of polarized gravitational cylindrical waves has
received a lot of attention in recent years \cite{Ashtekar:1996yk,
Ashtekar:1996bb, Angulo:2000ad, Gambini:1997yk, Dominguez:1999yk,
Varadarajan:1999aa, Cruz:1998ad, Romano:1996ep, Korotkin:1998ps}.
This is partly due to the fact that this system provides a
tractable, yet non-trivial, reduction of full general relativity
and hence is an ideal framework to explore several issues involved
in the quantization of gravity. Some intriguing phenomena, related
to the existence of large quantum gravity effects, have been
discussed by studying precisely this model \cite{Ashtekar:1996yk,
Ashtekar:1996bb, Angulo:2000ad,Gambini:1997yk, Dominguez:1999yk}.
It has also been argued that some manifestations of quantum
gravity, such as the smearing of light cones are, indeed, present
and can be understood in this simplified setting
\cite{Ashtekar:1996bb}.

One of the interesting points behind the obtained results is the
realization of the fact that the physical Hamiltonian is a
function of the free field Hamiltonian for a 2+1 dimensional,
axially symmetric, massless scalar field evolving in an auxiliary
Minkowski background \cite{Ashtekar:1996yk, Kuchar:1971xm,
Bicak:1997bb}. As we show in the first section of the paper, this
free Hamiltonian naturally appears when one linearizes the system,
thus suggesting that, in a precise sense, it can be considered as
the free part of an interacting model. However the full
interacting Hamiltonian is obtained by adding a very specific type
of terms to the free part, namely, just functions of it. Here we
plan to explore the consequences of this functional relation
between the two physically relevant Hamiltonians for the system
and explore how this affects the causal structure of quantum
spacetime. To this end we will pay attention to the smearing of
the light cones due to quantum gravity effects within the
framework of linearly polarized cylindrical waves. Instead of
considering the full information encoded in the metric tensor we
will concentrate on the causal structure provided by light cones.
An interesting, albeit somewhat indirect way, to look at this
structure is to study the commutators of field operators at
different spacetime points. These are the basic objects to discuss
the commutativity of observables and the microcausality of the
model; conventionally, a physical model should be such that
observables commute for space-like intervals. This has been
discussed in the standard perturbative quantum field theory
framework for simple examples such as scalar or fermion fields
(see e.g. \cite{peskin}). In fact, microcausality is one of the
key conditions to prove such important results as the
spin-statistics theorem \cite{peskin, pauli}.

Here we will use the commutator of the scalar field that describes
linearly polarized cylindrical waves as a way to get information
about the causal structure of quantum spacetime. As we show later
it is possible to give exact expressions for this commutator both
for the evolution provided by the free and the full physical
Hamiltonians. We will use these expressions to study in a
quantitative way the smearing of the light cones as a function of
the three-dimensional gravitational constant and explore some
physical issues, in particular the appearance of singularities as
a consequence of having a Hamiltonian bounded from above
\cite{Ashtekar:1994am, Varadarajan:1995ma}.

The rest of the paper is structured as follows. In Sec.
\ref{lingrav} we discuss how the free Hamiltonian is derived from
the linearized cylindrical wave model. Sec. \ref{evolutions} deals
with the classical and quantum dynamics of cylindrical
gravitational waves under the evolution provided both by the free
Hamiltonian and the physical Hamiltonian. We will pause here to
discuss and compare on a familiar example (the harmonic
oscillator) the main features of the time evolution defined by
functionally related Hamiltonians, both from the classical and the
quantum points of view. This will provide valuable insights for
the problem considered in this work. Sec. \ref{Microcausality} is
devoted to the study of microcausality. We will look at the main
features of the field commutators and study the smearing of the
light cones due to quantum gravitational effects. We end the paper
with a discussion of the main results and perspectives for future
work.


\section{\label{lingrav}Cylindrical Waves in Linearized Gravity}

Linearly polarized cylindrical waves in general relativity can be
described by the spacetime metric \cite{Ashtekar:1996bb,
Angulo:2000ad}:
\begin{equation}\label{met}
ds^2=e^{-\psi}ds_3^2+e^{\psi}dZ^2,\end{equation} where $Z\in
\mathbb{R}$ is the coordinate of the symmetry axis and $ds_3^2$ is
the three-metric
\begin{equation}
\label{3met} ds_3^2=-N^2
dt^2+e^{\gamma}(dR+N^{R}dt)^2+r^2d\theta^2.\end{equation} From
this three-dimensional point of view, $R\in \mathbb{R}^+$ and
$\theta\in S^1$ correspond to polar coordinates, $N^R$ is the
radial component of the shift vector and $N$ is the lapse
function. All metric functions ($\gamma$, $r$, $N$, and $N^R$)
depend only on the time and radial coordinates, $t$ and $R$.

Unless otherwise stated, we adopt a system of units such that
$c=\hbar=8G_3=1$, where $c$ is the speed of light, $\hbar$ is
Planck constant, and $G_3$ is the effective Newton constant per
unit length in the direction of the symmetry axis
\cite{Angulo:2000ad}. In these units, the gravitational action of
the system has the form \cite{Kuchar:1971xm,Romano:1996ep}
\begin{equation}\label{act}
S=\int_{t_1}^{t_2} dt \left[\int_0^\infty dR\,
(p_{\gamma}\dot{\gamma\;}+p_r\dot{r}+p_{\psi}\dot{\psi})-{\cal H}
\right],\end{equation} where the dot denotes the derivative with
respect to $t$, the $p$'s are the momenta canonically conjugate to
the metric variables and $\cal{H}$ is the total Hamiltonian
\begin{eqnarray}
\label{ham} {\cal H}=2\left(1-e^{-\gamma_{\infty}/2}\right)+
\int_0^\infty dR\left[NC+N^RC^R\right].\end{eqnarray} The first
term is a boundary contribution at infinity [$\gamma_{\infty}:=
\gamma(R\rightarrow\infty)$] and the second term is a linear
combination of the Hamiltonian constraint $C$ and the (radial)
diffeomorphisms constraint $C^R$:
\begin{eqnarray*}
C\;&=&e^{-\gamma/2}\left[2r^{\,\prime\prime}
-\gamma^{\prime}r^{\:\prime}-p_{\gamma}p_{r}+\frac{p_{\psi}^2}
{2r}+\frac{r(\psi^{\prime})^2}{2}\right] ,\nonumber\\
C^R&=&e^{-\gamma}\left(-2p^{\prime}_{\gamma}+
p_{\gamma}\gamma^{\prime}+p_rr^{\:\prime}+p_{\psi}\psi^{\prime}
\right).\end{eqnarray*} The prime denotes the derivative with
respect to $R$. The gauge freedom associated with these
constraints can be removed by imposing, respectively, the gauge
fixing conditions \cite{Ashtekar:1996bb,Angulo:2000ad}
\begin{eqnarray*}
\chi^R:=r-R=0,\hspace{.6cm}\chi\,:=p_{\gamma}=0.\end{eqnarray*}

On the other hand, the Lagrangian form of the action can be
obtained from the relations between momenta and time derivatives
of the metric provided by the Hamilton equations,
\begin{eqnarray*}
p_{\gamma}N &=& -e^{\gamma/2}\dot{r}+e^{-\gamma/2}N^Rr^{\:\prime},
\nonumber\\ p_{\psi} N &=& e^{\gamma/2}r\,\dot{\psi}-
e^{-\gamma/2}N^Rr\,\psi^{\prime},\nonumber\\ p_r N
&=&-e^{\gamma/2}\dot{\gamma\,}+2\left(N^Re^{-\gamma/2}\right)
^{\prime}.\end{eqnarray*}

From a three-dimensional perspective, the system describes an
axially symmetric model consisting of a scalar field $\psi$
coupled to gravity \cite{Ashtekar:1996bb}, the line element being
(\ref{3met}). A particular classical solution is a vanishing
scalar field in three-dimensional Minkowski spacetime or,
equivalently, Minkowski spacetime in four dimensions. In this
solution, $N=1$ and $r=R$, whereas the rest of metric fields and
momenta vanish (i.e.
$\psi=\gamma=N^R=p_{\psi}=p_{\gamma}=p_{r}=0$).

In this section, we will consider this solution as a background
and discuss first-order perturbations around it. In other words,
we will analyze the linearized theory of gravity around this
Minkowski spacetime, as it is usually done in the perturbative,
quantum field theory approach to gravity. In order to expand the
metric fields around the classical solution, let us call
\begin{eqnarray*}
r=R+\bar{r},\hspace{.6cm} N=1+\bar{N}.
\end{eqnarray*} Up to first-order terms in the fields, the expression of
the three-metric becomes
\begin{eqnarray*}
d\bar{s}_3^2&=&-(1+2\bar{N})
dt^2+2N^RdtdR+(1+\gamma)dR^2\nonumber\\
&&+(R^2+2R\bar{r})d\theta^2,\end{eqnarray*} while the
four-dimensional metric is given by
\begin{equation}\label{metlin}
d\bar{s}^2=(1-\psi)d\bar{s}_3^2+(1+\psi)dZ^2.\end{equation} Here,
it is understood that the product of $\psi$ with any other metric
field vanishes in the perturbative order considered. On the other
hand, regularity on the axis of symmetry imposes the following
conditions:
\begin{eqnarray}\label{reg}
\gamma(R=0)&=&0,\hspace{.6cm}N^R(R=0)=0,\nonumber\\
\bar{r}(R=0)&=&0,\hspace{.6cm}\bar{r}^{\,\prime}(R=0)=0.
\end{eqnarray}

In order to discuss the linearized gravitational system, we must
keep up to quadratic terms in the fields in the action
(\ref{act}). A straightforward calculation leads to the result
\begin{eqnarray*}
\bar{S}&=&\int_{t_1}^{t_2} dt\left[\int_0^\infty dR\, (p_{\gamma}
\dot{\gamma\;}+p_{\bar{r}}\dot{\bar{r}\;}+p_{\psi}\dot{\psi})
-\bar{\cal H} \right],\nonumber\\ \bar{\cal H}&=&\int_0^\infty dR
\left[
\frac{p_{\psi}^2}{2R}+\frac{R\,(\psi^{\prime})^2}{2}-p_{\gamma}
p_{\bar{r}}+\bar{N}\bar{C}+N^R\bar{C}^R\right]\nonumber\\ &&
+(2-\gamma_{\infty})\,\bar{r}^{\,\prime}(R\rightarrow\infty).
\end{eqnarray*}
Here, $p_{\bar{r}}:= p_r$ is the momentum canonically conjugate to
$\bar{r}$, and the linearized constraints are:
\begin{eqnarray*}
\bar{C}=2\bar{r}^{\,\prime\prime}-\gamma^{\prime},\hspace{.6cm}
\bar{C}^R= p_{\bar{r}}-2p_{\gamma}^{\prime}.\end{eqnarray*}

The diffeomorphisms gauge freedom can be fixed just like in the
cylindrical reduction of general relativity, namely, by demanding
that $r=R+\bar{r}$ coincides with the radial coordinate
\cite{Angulo:2000ad}. We thus impose the gauge fixing condition
$\bar{\chi}^R:=\bar{r}=0$. It is easily checked that the Poisson
brackets $\{\bar{\chi}^R,\bar{C}^R\}$ of this condition with the
linearized constraint do not vanish, so that the gauge fixing is
admissible. Dynamical consistency of the gauge fixing procedure
requires, in addition,
$$\dot{\;\bar{\chi}^R}=\{\bar{\chi}^R,\bar{\cal
H}\}=N^R-p_{\gamma}=0.$$ Hence, the shift must be chosen as
$N^R=p_{\gamma}$. Finally, the momentum conjugate to $\bar{r}$ is
fixed by solving the diffeomorphisms constraint:
$p_{\bar{r}}=2p_{\gamma}^{\prime}$. In this way, the canonical
pair $(\bar{r},p_{\bar{r}})$ is removed from the set of degrees of
freedom. The action of the resulting reduced model is
\begin{eqnarray*}
\bar{S}_1&=&\int_{t_1}^{t_2} dt \int_0^\infty dR\,
\left(p_{\gamma}\dot{\gamma\;}+p_{\psi}\dot{\psi}-\bar{N}
\bar{C}_1\right) \nonumber\\&-&\int_{t_1}^{t_2} dt \left[H_0
-p_{\gamma}^2(R\rightarrow\infty)+
p_{\gamma}^2(R=0)\right].\end{eqnarray*} Here,
$\bar{C}_1=-\gamma^{\prime}$ is the Hamiltonian constraint of the
reduced linearized system, and
\begin{equation}\label{ham0} H_0=\int_0^\infty dR\left[\frac{p_{\psi}^2}
{2R}+\frac{R\,(\psi^{\prime})^2}{2}\right].\end{equation}

Remarkably, the condition employed to eliminate the Hamiltonian
gauge freedom in full cylindrical gravity \cite{Angulo:2000ad} can
be used as well to fix the corresponding gauge in the linearized
theory. The gauge fixing $\bar{\chi}\,:=p_{\gamma}=0$ is
acceptable, because the Poisson brackets of $\bar{\chi}$ and
$\bar{C}_1$ differ from zero. In addition, consistency of the
chosen gauge demands the vanishing of
\begin{eqnarray*}
\dot{\bar{\chi}\,}=\left\{\bar{\chi}\; ,\;H_0+\!\int_0^\infty
\!\!dR\,
\bar{N}\bar{C}_1\right\}=-\bar{N}^{\prime}.\end{eqnarray*}
Therefore, $\bar{N}$ has to be independent of the radial
coordinate. Actually, we can set $\bar{N}=0$ by demanding that the
total lapse equals the unity at spatial infinity. On the other
hand, taking into account the regularity condition (\ref{reg}),
the solution to the Hamiltonian constraint is simply $\gamma=0$.
This allows us to remove the canonical pair $(\gamma, p_{\gamma})$
from the system and arrive at a constraint-free model in
linearized gravity.

The degrees of freedom of this system are the field $\psi$ and its
momentum. The reduced action is
\begin{eqnarray*}
\bar{S}_2=\int_{t_1}^{t_2} dt\left[- H_0+ \int_0^\infty\! dR\;
p_{\psi}\dot{\psi}\right].\end{eqnarray*} Note that $H_0$, given
in Eq. (\ref{ham0}), is the Hamiltonian of a massless scalar field
with axial symmetry in three-dimensions. Furthermore, in the gauge
that we have selected, the three-dimensional metric of the
linearized gravitational theory is exactly that of Minkowski
spacetime and contains no physical degrees of freedom. The scalar
field $\psi$ determines the norm of the Killing vector
$\partial_{Z}$, and appears in the four-dimensional metric of the
gauge-fixed, linearized model in the form (\ref{metlin}), but with
$d\bar{s}_3^2$ substituted by the flat metric
\begin{equation}
\label{mink} \left(d\bar{s}_3^2\right)_{f}:=
-dT^2+dR^2+R^2d\theta^2,\end{equation} where we have renamed $T$
the time coordinate of the reduced system.

Summarizing, the perturbative description provided in linearized
gravity for cylindrical waves with linear polarization around
four-dimensional Minkowski spacetime is equivalent to a massless
scalar field with axial symmetry in a three-dimensional flat
background. The dynamics of this field is dictated by the free
Hamiltonian $H_0$, which generates the evolution in the
Minkowskian time $T$.

It is worth noticing that the action and metric of the gauge-fixed
model in linearized gravity reproduce in fact the results that one
would obtain from the gauge-fixed model in full cylindrical
gravity by working just in the first perturbative order, i.e., by
keeping in the action and metric, respectively, up to quadratic
and linear terms in the field $\psi$ and its momentum. In this
sense, the gauge fixing and linearization procedures commute.


\section{\label{evolutions}Time Coordinates and Evolution for
Cylindrical Waves}

\subsection{\label{Functionally}Systems with Functionally Related
Hamiltonians}

One of the significant features of gravitational cylindrical waves
is the existence of two distinct, physically relevant Hamiltonians
(or equivalently two distinct time coordinates) to define both the
classical and quantum evolution. One is the Hamiltonian $H_0$
[given in Eq. (\ref{ham0})] that generates the dynamics in the
linearized gravitational theory; the other is the Hamiltonian $H$
that provides the energy per unit length along the symmetry axis
in general relativity \cite{Ashtekar:1994am, Varadarajan:1995ma,
Ashtekar:1996bb}. In fact, they are functionally dependent, since
$H=2(1-e^{-H_0/2})$. In order to gain insight into the relation
that can be established between the dynamics associated with these
two different Hamiltonians, we open this section by discussing a
similar situation in a very simple example provided by the
harmonic oscillator.

The usual description of the harmonic oscillator in a phase space
coordinatized by $(x_0,p_0)$ comes from its standard Hamiltonian
$h_0(x_0,p_0)=(p^2_0+\omega^2x^2_0)/2$. The dynamics is given by
the Hamilton equations
\begin{eqnarray*}
\frac{dx_0}{dT}=p_0\,, \quad \frac{dp_0}{dT}=-w^2x_0\,.
\end{eqnarray*}
The general solution can be written as
\begin{eqnarray*}
x_0(T)&=&\frac{1}{\sqrt{2\omega}}\left(ae^{-i\omega T}+a^\dagger
e^{i\omega T}\right),\\
p_0(T)&=&\frac{-i\omega}{\sqrt{2\omega}}\left(ae^{-i\omega
T}-a^\dagger e^{i\omega T}\right),
\end{eqnarray*}
where $a$ and its complex conjugate $a^\dagger$ are fixed by the
initial conditions.

Consider next a phase space $(x,p)$ with Hamiltonian $h=F(h_0)$,
i.e. a function of the standard Hamiltonian for the harmonic
oscillator. For instance, the case $F(h_o)=h_0^2$ arises in the
context of quantum optics, in relation with the propagation of
light in non-linear Kerr media \cite{Leonski:1998km}. The
equations of motion now read
\begin{eqnarray*}
\frac{dx}{dt}&=&\{x,F(h_0)\}=F'(h_0)p\,,\\ \frac{dp}{dt}&=&
\{p,F(h_0)\}=-\omega^2F'(h_0)x\,,
\end{eqnarray*}
where $F'$ denotes the derivative of $F$ with respect to its
argument. These new equations can be easily solved by means of a
change of time. Specifically, making use of the time independence
of the Hamiltonian $h_0$ ($h_0=w a^\dagger a$ on solutions to the
equations of motion), we can introduce the new time parameter
$T(t)=F'(h_0)t$. Then, the functions $x(t)=x_0[T(t)]$,
$p(t)=p_0[T(t)]$ and the new time $T$ serve us to transform the
Hamilton equations into the standard ones corresponding to the
harmonic oscillator. So, the classical solutions are
\begin{eqnarray*}
x(t)\!=\!x_0[T(t)]\!=\!\frac{1}{\sqrt{2\omega}}\!\left[ae^{\!
-i\omega F'(\omega a^\dagger\!a)t}\!+\!a^\dagger e^{i\omega
F'(\omega a^\dagger\!a)t}\right]\!,\\
p(t)\!=\!p_0[T(t)]\!=\!\frac{-i\omega}{\sqrt{2\omega}}\!
\left[ae^{\!-i\omega F'(\omega a^\dagger\!a)t}\!-a^\dagger
e^{i\omega F'(\omega a^\dagger\!a)t}\right]\!.
\end{eqnarray*}
What we find is an energy dependent redefinition of time that
induces a different time change for each solution to the equations
of motion.

The situation in the quantum theory is quite different. The reason
lies on the fact than in quantum mechanics a physical state does
not need to have a definite energy. Then, a ``change of time'' of
the form $\hat{T}=F^{\prime}(\hat{h}_0)t$ has non-trivial
consequences for the dynamics.

The usual quantum theory for the harmonic oscillator can be
described by introducing a Fock space with creation and
annihilation operators $a^\dagger$, $a$. Every initial state can
be expressed as $|\phi(0)\rangle=\sum_{n=0}^\infty c_n|n\rangle$,
where $c_n$ are Fourier coefficients (with the convenient
normalization) and $|n\rangle$ are energy eigenvectors (that is,
$\hat{h}_0|n\rangle =nw|n\rangle$). In the Schr\"odinger picture,
evolving with $\hat{h}_0$, we find
\begin{eqnarray*}
|\phi_0(T)\rangle&=&e^{-i\hat{h}_0T}|\phi(0)\rangle=\sum_{n=0}^\infty
c_ne^{-inwT}|n\rangle\,.
\end{eqnarray*}
However, if the same state $|\phi(0)\rangle$ evolves in time
according to the evolution generated by $\hat{h}=F(\hat{h}_0)$, we
get\footnote{Notice that the operators $\hat{h}_0$ and
$\hat{h}=F(\hat{h}_0)$ act on the same Hilbert space. Moreover, if
we demand the function $F$ to have a unique absolute minimum at
$0$, the two Hamiltonians have also the same vacuum.}
\begin{eqnarray*}
|\phi_F(t)\rangle&=&e^{-i\hat{h}t}|\phi(0)\rangle=\sum_{n=0}^\infty
c_ne^{-i F(nw)t}|n\rangle\,.
\end{eqnarray*}
Hence we do not recover an analogous situation to that found in
the classical system by replacing (formally) the time $T$ in
$|\phi_0(T)\rangle$ with $\hat{T}=F^{\prime}(\hat{h}_0)t$, because
$|\phi_F(t)\rangle \neq |\phi_0(\hat{T}(t))\rangle$ except for
linear homogeneous functions $F$. Moreover, the properties of the
states $|\phi_0(T)\rangle$ and $|\phi_F(t)\rangle$ are quite
different. For example, if we consider the bounded Hamiltonian
$\hat{h}=F(\hat{h}_0)=1-e^{-\hat{h}_0}$ it is obvious that the
high energy contributions to $|\phi_F(t)\rangle$ are essentially
frozen in time with respect to the evolution of the low energy
ones, in the sense that the phase of the former type of
contributions remains practically coherent in time.

\subsection{\label{Gravitational} Functionally Related Hamiltonians
and Cylindrical Waves}

Let us turn now to the discussion of our gravitational system. We
will deal with the Einstein-Rosen family of linearly polarized
gravitational waves \cite{Einstein:1937er}. These waves display
the so-called whole cylindrical symmetry \cite{Melvin:1965mt},
namely, they correspond to topologically trivial spacetimes which
possess two linearly independent, commuting, spacelike, and
hypersurface orthogonal Killing vector fields. It is well known
\cite{Kuchar:1971xm,Romano:1996ep} that these spacetimes admit
coordinates $(T,R,\theta,Z)$ such that the metric is given by
\begin{eqnarray*}
ds^2=e^{\gamma-\psi}(-dT^2+dR^2) +e^{-\psi}R^2d\theta^2+e^\psi
dZ^2,
\end{eqnarray*}
where $\psi$ and $\gamma$ are functions of only $R$ and $T$. When
the Einstein equations are imposed, $\psi$ encodes the physical
degrees of freedom and satisfies the usual wave equation for an
axially symmetric massless scalar field in three-dimensions:
\begin{eqnarray*}
\partial_T^2\psi-\partial_R^2\psi-\frac{1}{R}\partial_R\psi=0\,.\nonumber
\end{eqnarray*}
The metric function $\gamma$ can be expressed in terms of this
field on the classical solutions
\cite{Ashtekar:1996bb,Angulo:2000ad}. One gets
\begin{eqnarray}
\gamma(R)&=&\frac{1}{2}\int^R_0 d\bar{R} \,\bar{R}\, \left[
(\partial_T\psi)^2+(\partial_{\bar{R}}\psi)^2 \right],\nonumber\\
\label{gaminf} \gamma_\infty&=&\frac{1}{2}\int^\infty_0 dR\, R\,
\left[ (\partial_T\psi)^2+(\partial_R\psi)^2\right].
\end{eqnarray}
Note that $\gamma(R)$ and $\gamma_{\infty}$ are the energy of the
scalar field in a ball of radius $R$ and in the whole of the
two-dimensional flat space, respectively. Furthermore,
$\gamma_\infty$ coincides with the Hamiltonian $H_0$ given in Eq.
(\ref{ham0}) \cite{Ashtekar:1996bb}.

Nevertheless, to reach a unit asymptotic timelike Killing vector
field in the actual four-dimensional spacetime, with respect to
which one can truly introduce a physical notion of energy (per
unit length) \cite{Ashtekar:1994am, Varadarajan:1995ma}, one must
make use of a different system of coordinates, namely
$(t,R,\theta,Z)$ where $T=e^{-\gamma_\infty/2}t$. In these new
coordinates the metric has the form
\begin{eqnarray*}
ds^2=e^{\gamma-\psi}(-e^{-\gamma_\infty}dt^2+dR^2)
+e^{-\psi}R^2d\theta^2+e^\psi dZ^2.
\end{eqnarray*}
Assuming as a boundary condition that the metric function $\psi$
falls off sufficiently fast as $R\rightarrow \infty$, the above
metric describes asymptotically flat spacetimes with a, generally
non-zero, deficit angle. In this asymptotic region $\partial_t$ is
a unit timelike vector. The Einstein field equations can be
obtained from a Hamiltonian action principle
\cite{Ashtekar:1994am, Varadarajan:1995ma, Romano:1996ep} where
the on-shell Hamiltonian is given in terms of that for the free
scalar field by
\begin{eqnarray}
H=E(H_0)=2(1-e^{-H_0/2}). \label{newham}
\end{eqnarray}
Owing to these reasons, in the following we will refer to $t$ as
the physical time and to $H$ as the physical Hamiltonian.

When we use the $T$-time and impose regularity at the origin $R=0$
\cite{Ashtekar:1996bb}, the classical solutions for the field
$\psi$ can be expanded in the form
\begin{eqnarray*}
\psi(R,T)=\int_0^\infty \!\!\frac{dk}{\sqrt{2}}\, J_0(Rk)
\left[A(k)e^{-ikT} +A^\dagger(k)e^{ikT}\right].\nonumber
\end{eqnarray*}
$A(k)$ and $A^\dagger(k)$ are fixed by the initial conditions and
are complex conjugate to each other, because $\psi$ and $J_0$ (the
zeroth-order Bessel function of the first kind) are real. From Eq.
(\ref{gaminf}), we then obtain
\begin{eqnarray*}
\gamma_{\infty}=H_0=\int_0^{\infty}dk\,kA^{\dagger}(k)A(k).
\end{eqnarray*}
Using this formula, we can express the field in the $t$-frame as
\begin{eqnarray*}
\psi_E(R,t)=\hspace{6.4cm}\nonumber\\ \int_0^\infty\!\!
\frac{dk}{\sqrt{2}}\,
J_0(Rk)\left[A(k)e^{-ikte^{-\gamma_\infty/2}}
\!+\!A^\dagger(k)e^{ikte^{-\gamma_\infty/2}}\right].\;\nonumber
\end{eqnarray*}
Notice that $\psi(R,0)=\psi_E(R,0)$.

In principle, the quantization of the field $\psi$ can be carried
out in a standard way. We can introduce a Fock space in which
$\hat{\psi}(R,0)$, the quantum counterpart of $\psi(R,0)$, is an
operator-valued distribution \cite{Reed:1975rs}. Its action is
determined by those of $\hat{A}(k)$ and $\hat{A}^\dagger(k)$, the
usual annihilation and creation operators, whose only
non-vanishing commutators are
\begin{equation}\label{comma}
\left[\hat{A}(k_1),\hat{A}^\dagger(k_2)\right]=\delta(k_1,k_2).
\end{equation}
Explicitly,
\begin{equation}
\hat{\psi}(R,0)\!=\!\hat{\psi}_E(R,0)\!=\!\!\!\int_0^\infty
\!\!\!\frac{dk}{\sqrt{2}}\,
J_0(Rk)\!\left[\hat{A}(k)\!+\!\hat{A}^\dagger(k)\right]\!.
\label{psi0}
\end{equation}
In the Schr\"odinger picture, operators do not evolve, and the
problem of the time evolution is transferred to the physical
states. We will return later to this issue. In the Heisenberg
picture, on the contrary, operators change in time and states
remain fixed. In this case, the value of the quantum field
$\hat{\psi}$ at any time can be obtained from its value at
$T=t=0$, evolving in one of the two times that we have at hand.
One is the physical time $t$, with associated Hamiltonian $H$
given in Eq. (\ref{newham}). The other is the time $T$ of the
auxiliary three-dimensional Minkowski spacetime, its Hamiltonian
being that of a massless scalar field, $H_0$. From our discussion
in Sec. II, this time can be identified with the perturbative time
(denoted also by $T$) that arises in the study of the
Einstein-Rosen waves in linearized gravity.

In the perturbative time $T$, the evolution is provided by the
unitary operator $\hat{U}_0(T)=\exp(-iT\hat{H}_0)$ where
\begin{equation}\label{qham0}
\hat{H}_0=\int^{\infty}_0dk\,k\,\hat{A}^{\dagger}(k)\hat{A}(k)
\end{equation}
is the quantum Hamiltonian of an axially symmetric scalar field
in three dimensions. Then
\begin{eqnarray}
\hspace*{-.3cm}\hat{\psi}(R,T)\!\!\!&=&\!
\hat{U}^\dagger_0(T)\Psi(R,0)\hat{U}_0(T)\nonumber\\ \label{hpsiT}
&=&\!\!\!\!\int_0^\infty\!\! \frac{dk}{\sqrt{2}}\,
J_0(Rk)\!\left[\hat{A}(k)e^{-ikT}\!+\!\hat{A}^\dagger(k)e^{ikT}\right]\!.
\hspace*{.3cm}\end{eqnarray} The quantization procedure in this
case is very simple: we have substituted the initial conditions
$A(k)$ and $A^\dagger(k)$ in the classical solution by the
corresponding quantum operators.

The situation changes when we choose the physical time $t$ as the
time parameter. The quantum Hamiltonian can be defined as
$\hat{H}=E(\hat{H}_0)=2(1-e^{-\hat{H}_0/2})$. We can then reach a
unitary evolution by means of $\hat{U}(t)=\exp(-it\hat{H})$. This
leads to the following time evolved operators
\begin{eqnarray}
\hat{A}_E(k,t)\!\!&:=&\!\!\hat{U}^\dagger(t)\hat{A}(k)\hat{U}(t)=
\exp\!\left[-itE(k)e^{-\hat{H}_0/2}\right]\!\hat{A}(k),\nonumber\\
\label{crean}
\hat{A}_E^\dagger(k,t)\!\!&=&\!\hat{A}^\dagger(k)\,\exp\!\left[itE(k)
e^{-\hat{H}_0/2}\right],
\end{eqnarray}
where $E(k)=2(1-e^{-k/2})$. It is important to realize that the
quantum evolution in the physical $t$-frame is not obtained by
changing $A(k)$, $A^\dagger(k)$, and $\gamma_\infty$ by their
direct quantum counterparts. In fact, by restoring the value of
the dimensionful constants $\hbar$ and $G_3$, we can write
$E(k)=(1-e^{-4\bar{G}k})/(4G_3)$ with $\bar{G}=\hbar G_3$, so that
\begin{eqnarray*}
\frac{t}{\hbar}E(k)=tk+o(\hbar),
\end{eqnarray*}
and we can expand $\hat{A}_E(k,t)$ and $\hat{A}_E^\dagger(k,t)$ in
powers of $\hbar$,
\begin{eqnarray*}
\hat{A}_E(k,t)&=&\exp\left(-itke^{-4\bar{G}\hat{H}_0}\right)
\hat{A}(k)+o(\hbar),\\ \hat{A}_E^\dagger(k,t)&=&\hat{A}^\dagger(k)
\exp\left(itke^{-4\bar{G}\hat{H}_0}\right)+o(\hbar).
\end{eqnarray*}
(Here, we have expressed $\hat{H}_0$ with dimensions of an inverse
length)

Setting again $8\bar{G}=1$, the expansion above clearly shows that
the quantum evolution of the creation and annihilation variables
in the physical time differs from the ``classical evolution" in
higher-order quantum corrections.

This unusual behavior can be partially corrected in the sense that
one can actually find an operator for $H$ such that the quantum
evolution in the $t$-time is similar to the classical one. In
fact, if one considers the normal ordered Hamiltonian
$\hat{H}_{nor}=:\hat{H}:$ and its associated unitary evolution
operator $\hat{U}_{nor}=\exp(-it\hat{H}_{nor})$, it is easy to
prove that
\begin{eqnarray*}
\hat{A}_{nor}(k,t)&=&\!\exp\left(-itk:e^{-\hat{H}_0/2}:\right)\hat{A}(k),
\\
\hat{A}^\dagger_{nor}(k,t)&=&\!\hat{A}^\dagger(k)
\exp\left(itk:e^{\hat{H}_0/2}:\right).
\end{eqnarray*}
Therefore, the quantum evolution in $t$ given by $\hat{H}_{nor}$
parallels that encountered in the classical picture.
Unfortunately, there is a severe problem with this quantum
dynamics that renders it physically unacceptable: the Hamiltonian
$\hat{H}_{nor}$ is unbounded both from above and below
\cite{Ashtekar:1996bb}.

Obviously, the different possibilities for the unitary time
evolution of the field $\hat{\psi}$ considered here are closely
related. Specifically, there exists a unitary mapping between both
types of evolution. This is a consequence of the fact that they
are both unitary and coincide with the identity in the same
Hilbert space at $T=t=0$. So, if we denote the operators evolved
from $\hat{X}(0)$ in the times $T$ and $t$ by $\hat{X}(T)=
 \hat{U}_0^\dagger(T)\hat{X}(0)\hat{U}_0(T)$ and
$\hat{X}_E(t)=\hat{U}^\dagger(t)\hat{X}(0)\hat{U}(t)$,
respectively, we obtain
\begin{eqnarray*}
\hat{X}_E(t)=\hat{U}^\dagger (t)\hat{U}_0(T) \hat{X}(T)
\hat{U}^\dagger_0 (T)\hat{U}(t)\,.
\end{eqnarray*}
Then, we can go from one evolution to the other by means of the
unitary operator $\hat{U}^\dagger_0 (T)\hat{U}(t)$.

Let us close this section with a few comments about the Fock space
on which the field $\hat{\psi}$ acts. In the Heisenberg picture
the states do not depend on time. They are constructed by
successive actions of the time independent creation operators on
the vacuum of the theory. It is important to point out that both
$\hat{H}_0$ and $\hat{H}$ act on the same Hilbert space and have
the same vacuum, that will be referred to as $|0\rangle$.
Explicitly, given any square integrable complex function
$\phi_n(k_1,\dots,k_n)$ (with the convenient normalization) we can
write a $n$-particle state in the form
\[
|\phi_n\rangle\!=\!\!\!\int_0^\infty
\hspace{-.35cm}dk_1\hspace{-.1cm}
\cdots\hspace{-.1cm}\int_0^\infty \hspace{-.35cm}dk_n
\phi_n(k_1,\dots,k_n)\hat{A}^\dagger(k_1)\hspace{-.05cm}\cdots
\hat{A}^\dagger(k_n)|0\rangle.
\]
According to the usual interpretation of quantum mechanics, the
measurable physical quantities correspond to expectation values of
observables. We can go over to the Schr\"odinger picture by
assigning the time evolution to the quantum states
\begin{eqnarray*}
X(t;\phi)\!=\!\langle \phi(0)|\hat{U}^\dagger\!
(t)\hat{X}(0)\hat{U}\!(t)|\phi(0)\rangle\!=\!\!\langle
\phi(t)|\hat{X}(0)|\phi(t)\rangle.
\end{eqnarray*}
Defining $|\phi_E(t)\rangle=\hat{U}(t)|\phi(0)\rangle$,
$|\phi(T)\rangle=\hat{U}_0(T)|\phi(0)\rangle$, and noticing that
the $U$-operators satisfy
\begin{eqnarray*}
i\partial_t\hat{U}(t)=\hat{H} \hat{U}(t),\quad\quad
i\partial_T\hat{U}_0(T)=\hat{H}_0 \hat{U}_0(T)\,,
\end{eqnarray*}
it is straightforward to see that the evolved states are solutions
to the Schr\"odinger equations
\begin{eqnarray*}
i\partial_t|\phi_E(t)\rangle=\hat{H} |\phi_E(t)\rangle ,\quad\quad
i\partial_T|\phi(T)\rangle=\hat{H}_0 |\phi(T)\rangle \,.
\end{eqnarray*}
As in the Heisenberg picture, the unitary operator $\hat{U}(t)
\hat{U}^\dagger_0(T)$ provides the bridge between the two kinds of
quantum evolution,
\begin{eqnarray*}
|\phi_E(t)\rangle=\hat{U}(t)
\hat{U}^\dagger_0(T)|\phi(T)\rangle\,.
\end{eqnarray*}
Particularizing the above results to the case of $n$-particle
states we readily get that, for the $T$-time,
\begin{eqnarray*}
|\phi_n(T)\rangle=\hat{U}_0(T)|\phi_n\rangle=\hspace{4.4cm}\\
\int_0^\infty \hspace{-.35cm}dk_1\hspace{-.1cm}
\cdots\hspace{-.15cm}\int_0^\infty \hspace{-.35cm}dk_n
e^{\!-ik_{tot}T}\!\phi_n(k_1,...,k_n)\hat{A}^\dagger\!(k_1)
\hspace{-.05cm}\cdots\hat{A}^\dagger\!(k_n)|0\rangle
\end{eqnarray*}
where $k_{tot}=\sum_{j=1}^nk_j$. Notice that $|\phi_n(T)\rangle$
is a superposition of eigenvectors of $\hat{H}_0$ with eigenvalues
equal to $k_{tot}$. On the other hand, if we evolve the states
with $\hat{U}(t)$, it is not difficult to check that
\begin{eqnarray}
|\phi_{E,n}(t)\rangle=\hat{U}(t)|\phi_n\rangle=&&\nonumber\\
\int_0^\infty \hspace{-.35cm}dk_1\hspace{-.1cm}
\cdots\hspace{-.2cm}\int_0^\infty \hspace{-.35cm}dk_n
e^{\!-iE(k_{tot})t}&&\hspace*{-.4cm}\phi_n(k_1,...,k_n)
\hat{A}^\dagger\!(k_1)\!\cdots\hspace{-.1cm}
\hat{A}^\dagger\!(k_n)|0\rangle.\nonumber
\end{eqnarray}
In other words, $|\phi_{E,n}(t)\rangle$ is a superposition of
eigenvectors of $\hat{H}$, each of them with energy $E(k_{tot})$.
Finally, it is worth pointing out that the $\hat{H}$-energy is not
additive:
\[
E(k_{tot})\!=\!2\!-\!2\,e^{\!-(\sum_1^n k_i)/2}\!\!\neq
\!\!\sum_{i=1}^n\!E(k_i)\!=\! 2n-2\!\sum_{i=1}^n\!e^{\!-k_i/2}\!.
\]
This property is directly related to the existence of an upper
bound for the physical Hamiltonian.


\section{\label{Microcausality} Microcausality}

\subsection{\label{H0} Free Hamiltonian}

Microcausality plays a crucial role in perturbative quantum field
theory; in fact it is a crucial ingredient in such important
issues as the spin-statistics theorem \cite{pauli}. The point of
view that we will develop in this section is the idea that field
commutators of the scalar field that encodes the physical
information in linearly polarized cylindrical waves can be used as
an alternative to the metric operator to extract physical
information about quantum spacetime. Some relevant information may
be lost, but the availability of explicit exact expressions for
commutators [even under the evolution given by the Hamiltonian
(\ref{newham})] opens up the possibility of getting precise
information about the quantum causal structure of spacetime. In
particular we will see how the light cones get smeared by quantum
corrections in a precise and quantitative way.

As is well known, the study of causality in perturbative quantum
field theory requires the consideration of measurements of
observables at different spacetime points and their mutual
influence. The key question is whether measurements taken at
spatial separations commute or not. The relevant commutators can
be seen to be proportional to those of the quantum fields at two
different spacetime points $x$ and $y$. For a free massless scalar
field described by the standard Lagrangian, this commutator is a
(c-number) function of $x$ and $y$ that is exactly zero when $x-y$
is space-like \cite{peskin}. This means that observables at points
separated by spatial intervals commute. Something similar happens
for fermions described by the usual Lagrangians if, instead of
commutators, one takes anticommutators of the fields (observables
for fermion fields can be written as even powers of them and the
relevant commutators can be written in terms of anticommutators of
the basic fields) \cite{peskin}.

In the case that we are considering in this work the only physical
local degree of freedom that we have is given by the scalar field
$\psi$. What we will do in the following is discussing
microcausality by considering the different time evolutions
introduced in the first part of the paper.

We start by computing $[\hat{\psi}(R_1,T_1),\hat{\psi}(R_2,T_2)]$
for the field operators (\ref{hpsiT}) obtained in the Heisenberg
picture by evolving the field at time $T=0$ with the Hamiltonian
$\hat{H}_0$ given in Eq. (\ref{qham0}). It is straightforward to
get (see \cite{Allen:1987al} for a somewhat related computation)
\begin{eqnarray}
[\hat{\psi}(R_1,T_1),\hat{\psi}(R_2,T_2)]=\hspace{4.cm}\nonumber\\
\label{s2-004} i\int_0^\infty\!\!\!dk J_0(R_1k)J_0(R_2k)
\sin[(T_2-T_1) k ].\hspace*{.3cm}
\end{eqnarray}
Let us discuss the main features of this commutation function,
that we will refer to in the following as the $H_0$-commutator. To
begin with we can easily see that, as it happens for the familiar
free field theories, the commutator (\ref{s2-004}) is a c-number,
i.e. it is proportional to the identity in the Fock space. In
addition, for $R_1$ fixed, let us call regions I, II, and III the
regions of the $(R_2,T_2-T_1)$ plane defined, respectively, by
$0<|T_2-T_1|<|R_2-R_1|$, $|R_2-R_1|<|T_2-T_1|<R_2+R_1$, and
$R_1+R_2<|T_2-T_1|$. Then, it can be shown that the
$H_0$-commutator vanishes in region I, whereas in region II it can
be written as \cite{Gradshteyn:1994gr}
\begin{eqnarray}
[\hat{\psi}(R_1,T_1),\hat{\psi}(R_2,T_2)]=\hspace*{4.4cm}&&
\nonumber \\ \label{propII} \frac{i}{\pi}\frac{1}{\sqrt{ R_1
R_2}}\,K\left(\sqrt{\frac{(T_2\!-\!T_1)^2\!-\!(R_2\!-\!R_1)^2}
{4R_1R_2}}\right).\hspace*{1.cm}&&
\end{eqnarray}
Finally, its value in region III is \cite{Gradshteyn:1994gr}
\begin{eqnarray}[\hat{\psi}(R_1,T_1),\hat{\psi}(R_2,T_2)]=\hspace*{.5cm}
&&\nonumber\\
\frac{2i}{\!\sqrt{\!\pi^2[(T_2\!-\!\!T_1)^2\!\!-\!(R_2\!-\!
\!R_1)^2]}}&&\hspace*{-.4cm}
K\!\!\left(\!\!\sqrt{\!\frac{4R_1R_2}{(T_2\!-\!\!T_1)^2\!\!-
\!(R_2\!-\!\!R_1)^2}}\right)\!.\nonumber
\end{eqnarray}
Here, $K$ denotes the complete elliptic integral of the first
kind, $K(k):=\int_0^{\pi/2}\!d\theta/\sqrt{1-k^2 \sin^2\theta}$
(alternatively, the integral in Eq. (\ref{s2-004}) can be written
in terms of the associated Legendre functions $P_{-1/2}$ and
$Q_{-1/2}$ \cite{Gradshteyn:1994gr}).
\begin{figure}
\includegraphics[width=8.5cm]{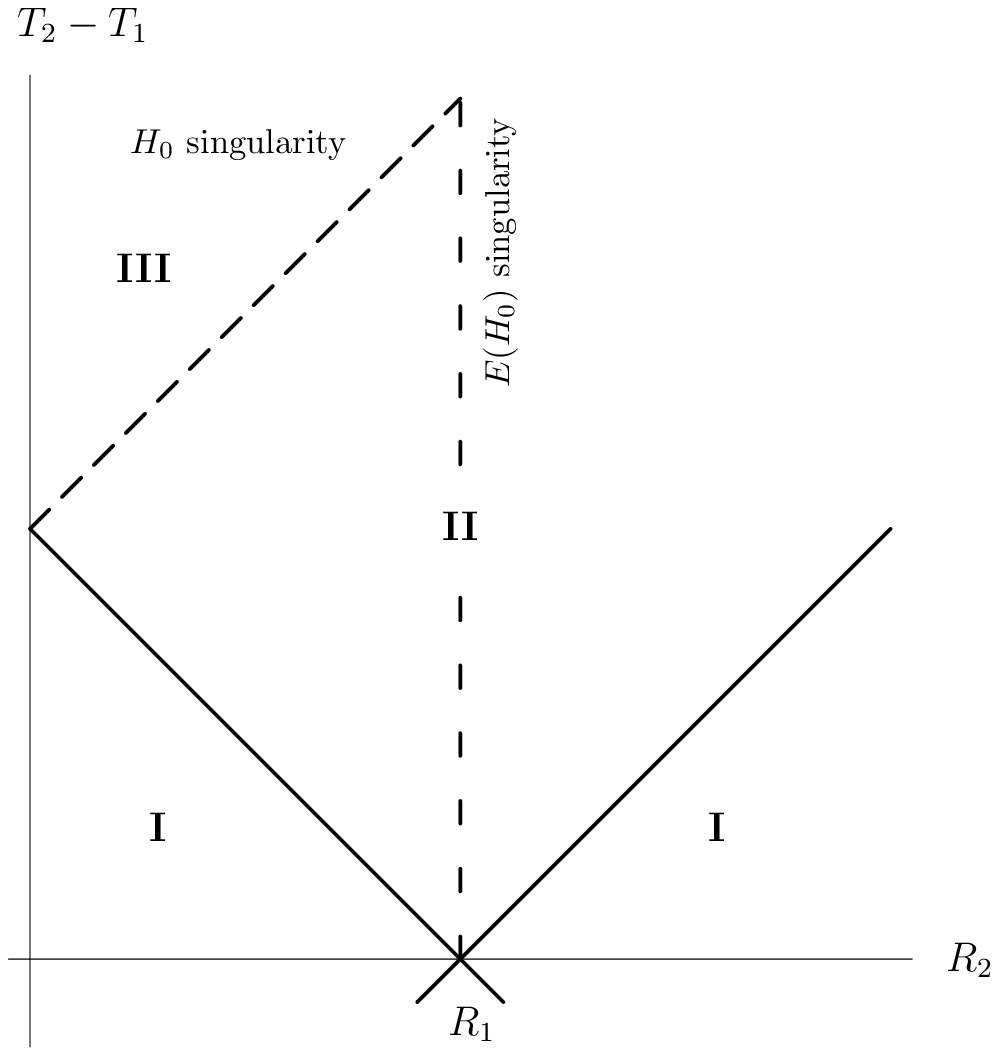}
\caption{Regions in the $(R_2,T_2-T_1)$ plane for the
$H_0$-commu\-tator. Region I corresponds to
$0<|T_2\!-\!T_1|<|R_2-R_1|$, region II to
$|R_2-R_1|<|T_2-T_1|<R_2+R_1$, and region III to
$R_2+R_1<|T_2-T_1|$. The singularity of the $H_0$-commu\-tation
function lies in the boundary between regions II and III, whereas
the singularity for the $E(H_0)$-commutator appears for
$R_1=R_2$.} \label{fig2:regions}
\end{figure}

\begin{figure}
\includegraphics*[width=8.5cm]{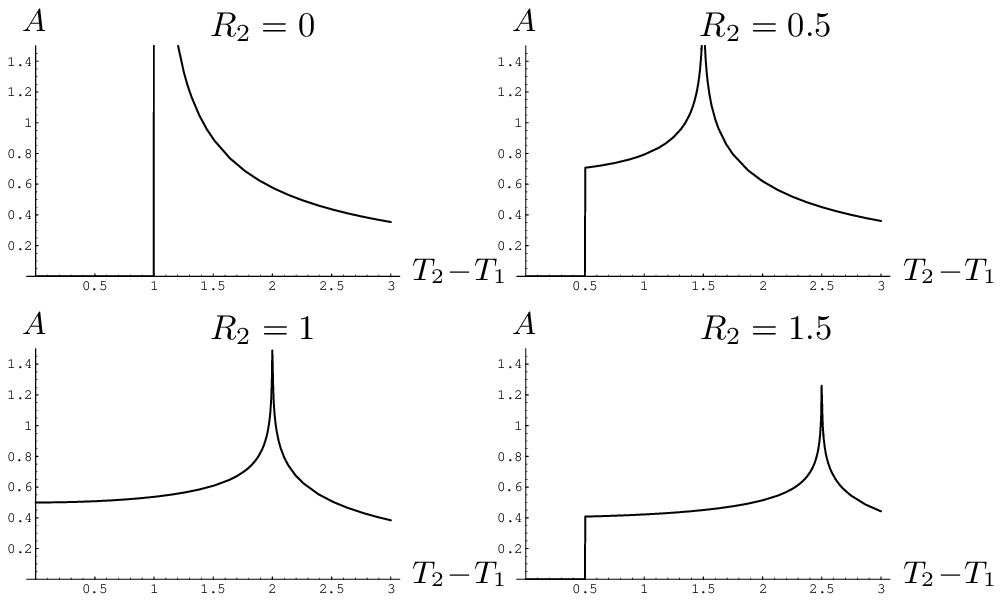}
\caption{Absolute value $A$ of the commutator of the scalar field
$[\hat{\psi}(R_1,T_1),\hat{\psi}(R_2,T_2)]$ for $R_1=1$ and
several values of $R_2$ as a function of
$T_2-T_1$.}\label{fig:freeprop2}
\end{figure}

\begin{figure}
\includegraphics[width=8.5cm]{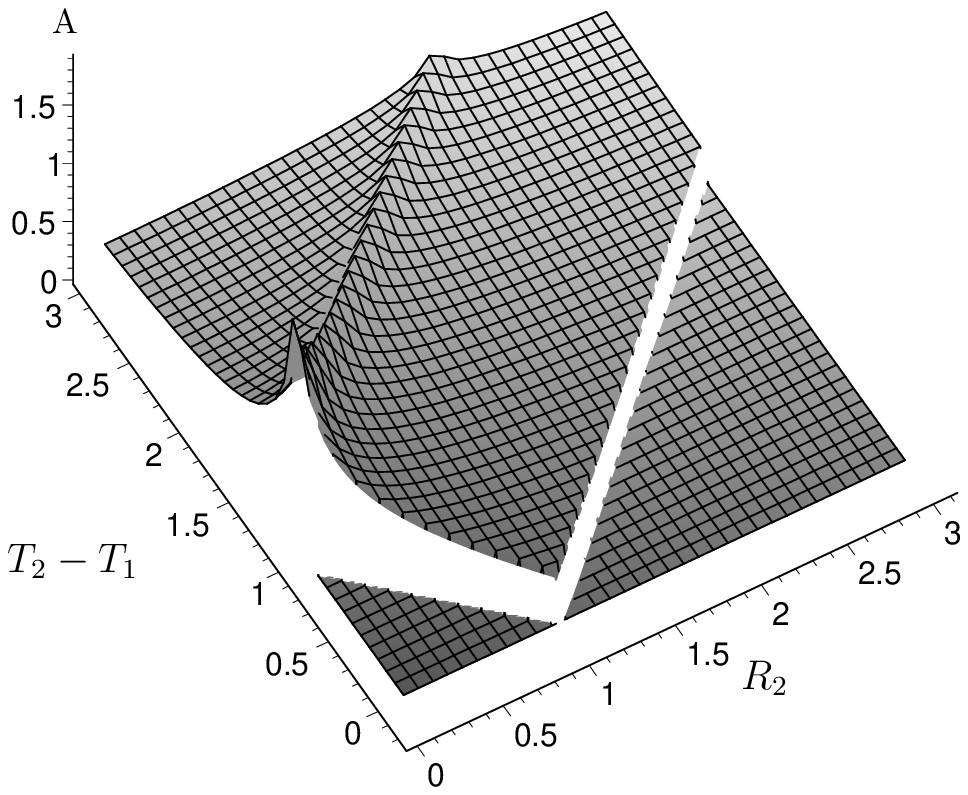}
\caption{Absolute value $A$ of the commutator of the scalar field
$[\hat{\psi}(R_1,T_1),\hat{\psi}(R_2,T_2)]$ for $R_1=1$ as a
function of $T_2-T_1$ and $R_2$. The light cone structure and the
singularity at $T_2-T_1=R_2+R_1$ can be readily seen. The plot for
negative values of $T_2-T_1$ is obtained by reflecting with
respect to the $R_2$ axis.} \label{fig:freeprop}
\end{figure}

Several plots of the absolute value of this function (in fact, of
its imaginary part) for fixed values of $R_1$ and $R_2$ are shown
in Fig. \ref{fig:freeprop2}, and a three-dimensional plot for a
fixed value of $R_1$ can be found in Fig. \ref{fig:freeprop}.
Some interesting properties can be read off from the previous
expressions. First of all we see that the commutator is in fact
identically zero in the region labeled I; outside it differs from
zero. Second we see that it is singular in the line
$T_2-T_1=R_2+R_1$. It is easy to check that the singularity is
logarithmic by using the explicit form of the propagator
(\ref{propII}) and expanding around $T_2-T_1=R_2+R_1$. This
singularity can be understood as a consequence of the presence of
the symmetry axis; this is supported by the fact that the
$H_0$-commutation function satisfies
\begin{eqnarray}
&&\left(\!\partial_T^2-\frac{1}{R}\partial_R(R\partial_R)\!\right)\!\cdot
[\hat{\psi}(R,T),\hat{\psi}(R_2,T_2)]=0\,,\nonumber\\
&&\left(\!\partial_T^2-\frac{1}{R}\partial_R(R\partial_R)\!\right)\!\cdot
\theta(T-T_2)[\hat{\psi}(R,T),\hat{\psi}(R_2,T_2)]=\nonumber\\
&&\hspace{3.cm}\frac{i}{R_2}\delta^{(2)}\!(R-R_2,T-T_2)\,,\nonumber
\end{eqnarray}
which shows that the full commutator is an axially symmetric
solution to the 2+1 dimensional wave equation and the future part
of it (obtained by multiplying by a step function) is an axially
symmetric Green function for the same equation. It is important to
point out that even though there is a Minkowskian background
metric, the presence of a center of symmetry breaks Lorentz
invariance; in fact, the only symmetries of the 2+1 axially
symmetric wave equation
$$
\left(\!\partial_T^2-\frac{1}{R}\partial_R(R\partial_R)\!\right)
\Phi(T,R)=0
$$
that transform solutions $\Phi(T,R)$ into solutions\footnote{In
addition to the ones coming from its linear and homogeneous
character.} are

i) Translations
\begin{eqnarray*}
\Phi(T,R)\mapsto \Phi(T-\epsilon,R) \quad (\epsilon\in\mathbb{R}),
\end{eqnarray*}

ii) Dilatations
\begin{eqnarray*}
\Phi(T,R)\mapsto \Phi(e^{\epsilon}T,e^{\epsilon}R)
\quad(\epsilon\in \mathbb{R}),
\end{eqnarray*}

iii) Inversions
\begin{eqnarray*}
\Phi(T,R)\!\mapsto
\frac{\Phi[\tau_\epsilon(T,\!R),\rho_\epsilon(T,\!R)]}
{\sqrt{1\!-\!2T\epsilon\!+\!(T^2\!-\!R^2)\epsilon^2}}\!\quad(\epsilon\textrm{
small enough})
\end{eqnarray*}
with
\begin{eqnarray*}
\tau_\epsilon(T,R)&=&\frac{T-\epsilon(T^2-R^2)}{1-2T\epsilon+(T^2-R^2)\epsilon^2}\,,\\
\rho_\epsilon(T,R)&=&\frac{R}{1-2T\epsilon+(T^2-R^2)\epsilon^2}\,.
\end{eqnarray*}
These can be obtained systematically by using the general theory
of symmetry groups for partial differential equations (see
\cite{olver}).

\subsection{\label{E(H0)} $E(H_0)$  Hamiltonian}

Let us consider now the commutator of the field operators obtained
in the Heisenberg picture with the quantum Hamiltonian
$\hat{H}:=E(\hat{H}_0)=(1-e^{-4\bar{G} \hat{H}_0})/(4G_3)$. Here
$\bar{G}=G_3\hbar$, and we have restored the values of the Planck
constant and the three-dimensional gravitational constant (but
kept $c=1$) to have the possibility of discussing the
semiclassical limit. Recall that the Hamiltonian $\hat{H}_0$ is
given by Eq. (\ref{qham0}), where we choose the normalization of
the creation and annihilation operators so that the commutation
relation (\ref{comma}) is satisfied (with no $\hbar$ in the
r.h.s.). Notice that, with this convention, $\hat{A}(k)$,
$\hat{H}_0$, and $\hat{H}$ have formally the same dimensions as
$\bar{G}^{1/2}$ (or $k^{-1/2}$), $k$, and $G_3^{-1}$ (or $\hbar
k$), respectively. To distinguish the field operators evolved with
the new Hamiltonian $\hat{H}$, we will denote them by
$\hat{\psi}_E(R,t)$. In order to compute the commutator
$[\hat{\psi}_E(R_1,t_1),\hat{\psi}_E(R_2,t_2)]$, we make use of
the expressions (\ref{crean}) for the creation and annihilation
operators evolved in $t$, which can also be written as
\begin{eqnarray}
\hat{A}_E(k,t)\! &=&\!
\hat{A}(k)\exp\left[it\left(E(\hat{H}_0-k)-E(\hat{H}_0)\right)\right],
\nonumber\\ \label{crean2}
\hat{A}^\dagger_E(k,t)\!&=&\!\hat{A}^\dagger(k)\exp\left[it
\left(E(\hat{H}_0+k)-E(\hat{H}_0)\right)\right].\hspace*{.9cm}
\end{eqnarray}
Substituting then the relation
\begin{eqnarray*}
\hat{\psi}_E(R,t)=\sqrt{4\bar{G}}\int_0^\infty\!\!\!dk
\,J_0(Rk)\left[\hat{A}_E(k,t)+\hat{A}_E^\dagger(k,t)\right],
\end{eqnarray*}
we get the relevant field commutator
\begin{eqnarray*} \big[\hat{\psi}_E(R_1,t_1),
\hat{\psi}_E(R_2,t_2)\big]\!\!=\!\!\!
\int_0^\infty\!\!\!\!\!\!\!dk_1\!\!\int_0^\infty\!\!\!\!
\!\!\!dk_2J_0(R_1k_1)J_0(R_2k_2)\\ \times 4\bar{G}
\bigg\{\!\big[\hat{A}_E(k_1,\!t_1),\hat{A}_E(k_2,\!t_2)\big]\!+\!
\big[\hat{A}^\dagger_E(k_1,\!t_1),\hat{A}^\dagger_E(k_2,\!t_2)\big]
\\ +\big[\hat{A}_E(k_1,\!t_1),\hat{A}^\dagger_E(k_2,\!t_2)\big]\!+\!
\big[\hat{A}^\dagger_E(k_1,\!t_1),\hat{A}_E(k_2,\!t_2)\big]
\bigg\}.
\end{eqnarray*}
The commutators involving $\hat{A}_E(k,t)$ and
$\hat{A}^{\dagger}_E(k,t)$ can be found in Appendix \ref{A1}. Note
that, in contrast with the evolution given by $H_0$, for which the
commutation function is a c-number, the situation now is more
complicated because
$\big[\hat{\psi}_E(R_1,t_1),\hat{\psi}_E(R_2,t_2)\big]$ is an
operator, as it happens in interacting theories. We will analyze
two types of matrix elements for it, namely, the expectation value
on the vacuum and on one-particle states. In addition, we will
briefly comment on the expectation value on coherent states.
\bigskip

\noindent$\bullet$ Vacuum expectation value.

\begin{eqnarray}
\langle 0|\,
[\hat{\psi}_E(R_1,t_1),\hat{\psi}_E(R_2,t_2)]\,|0\rangle=\hspace*{3.3cm}&&
\nonumber\\ i8\bar{G}\!\!\int_0^{\infty}\!\!\!\!\!\!dk\,J_0(R_1
k)J_0(R_2
k)\sin\left[\frac{t_2-t_1}{4\bar{G}}(1\!-\!e^{-4\bar{G}k})\right]\!.
\hspace*{.4cm}&&\label{s2-005}
\end{eqnarray}
In the following we will discuss the differences and similarities
of this matrix element and the $H_0$-commutation function.

\begin{figure}
\includegraphics[width=8.5cm]{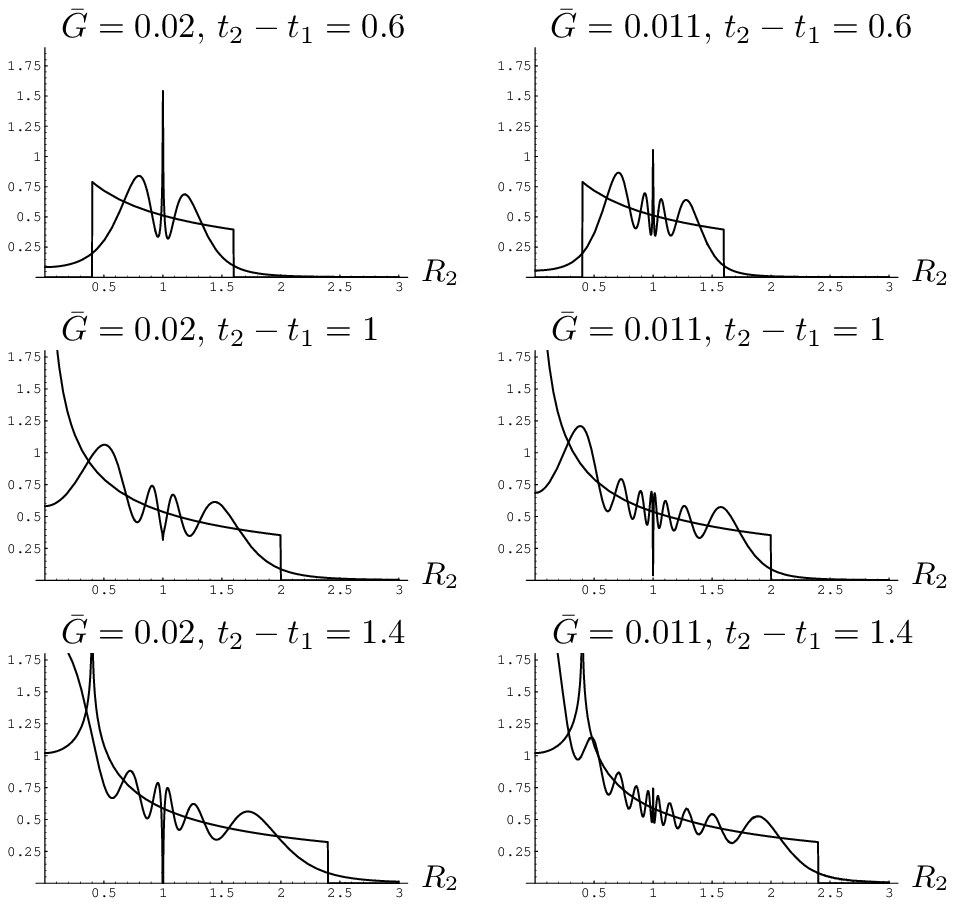}
\caption{Comparison between the absolute values (over $8\bar{G}$)
of the $H_0$-commutator and the vacuum expectation value of the
$E(H_0)$-commutator for two different values of $\bar{G}$, plotted
for $R_1=1$.}\label{fig4:compare}
\end{figure}
We point out that the factor $\sin[(T_2-T_1) k]$ that appears in
the integrand of the $H_0$-commutator is substituted now by
$\sin[(t_2-t_1)(1-e^{-4\bar{G}k})/(4\bar{G})]$. They coincide for
$k\rightarrow0$, but the former of these functions oscillates for
all values of $k$ whereas the latter approaches a constant value
when $k\rightarrow\infty$. This changes the convergence properties
of the integral. In particular it is straightforward to see that
the integral (\ref{s2-005}) diverges whenever $R_1=R_2$ (except if
$\sin\left[(t_2-t_1)/(4\bar{G})\right]=0$) but converges
otherwise. Therefore, the vacuum expectation value has a
singularity structure that differs from the one given by Eq.
(\ref{s2-004}). This has some interesting physical consequences.
First we see that the singularity that originates in the axis in
the $H_0$ case is not present when the evolution is dictated by
$E(H_0)$; this can be interpreted as a blurring of the axis due to
quantum corrections. Second we see that a completely different
kind of singularity pops up when the evolution is generated by
$E(H_0)$. From a mathematical point of view its origin is clearly
related to the fact that the energy is bounded from above and,
hence, for large values of $k$ the integrand is just a product of
two $J_0$ Bessel functions (which give a divergent integral if
their arguments coincide). Physically, the emergence of the
singularity can be understood in an intuitive manner by writing a
state as a superposition of vectors of the form
$\hat{A}^{\dagger}(k)|0\rangle$,
\[
\hat{\psi}_E(R_1,t_1)|0\rangle\!=\!\sqrt{4\bar{G}}\!
\int_0^{\infty}\!\!\!\!\!\!dk\,J_0(R_1k)\,
e^{\frac{it_1}{4\bar{G}}(1-e^{-4\bar{G}k})}
\hat{A}^{\dagger}(k)|0\rangle,
\]
and projecting onto $\hat{\psi}_E(R_2,t_2)|0\rangle$. The
$e^{it_{j}(1-e^{-4\bar{G}k})/(4\bar{G})}$ factor (for $j=1$ or 2)
goes to a phase that depends only on $t_j$ as
$k\rightarrow\infty$. So, if $R_1=R_2$, the coefficients of the
linear superposition defining, respectively,
$\hat{\psi}_E(R,t_1)|0\rangle$ and $\hat{\psi}_E(R,t_2)|0\rangle$
differ only by a constant phase for large values of $k$. This
means that, in the sector of large $k$, these two states have
coherent phases and therefore a constructive interference. Since
each of them has an infinite norm, their scalar product diverges.
A similar effect can be found in the quantum dynamics of a free
particle with an energy given by a function $E(k)$ with
$dE(k)/dk\rightarrow 0$ as $k\rightarrow\infty$. If one builds a
wave packet as
$$
\Phi(t,x)=\int_0^\infty dk\, C(k) e^{-i[tE(k)+kx]},
$$
with $C(k)$ peaked around a large value of $k$, $k_0\gg 1$, the
group velocity becomes almost zero and $\Phi(t,x)$ stays
essentially the same at every $x$ for long periods of time.

The integral in Eq. (\ref{s2-005}) can be written explicitly as a
convergent series (when $R_1\neq R_2$) by expanding the sine
function as a power series of $e^{-4\bar{G}k}$ and computing the
resulting integrals involving two Bessel functions and an
exponential \cite{Gradshteyn:1994gr}
\begin{eqnarray}
\langle 0|\,
[\hat{\psi}_E(R_1,t_1),\hat{\psi}_E(R_2,t_2)]\,|0\rangle=
\hspace*{3.3cm}&&\nonumber
 \label{s2-006}\\
\frac{i8\bar{G}}{\pi\sqrt{R_1R_2}}\left\{\!\sin(\Delta
t)\!\!\sum_{n=0}^{\infty} \!\frac{\!(-1)^n(\Delta t)^{2n}\!\!}
{(2n)!}Q_{-1/2}\left[\sigma_{2n}(R_1,R_2)\right]\right.
&&\nonumber\\ -\left.\!\cos(\Delta t)\!\sum_{n=0}^{\infty}
\!\frac{\!(-1)^n(\Delta t)^{2n+1}\!\!\!}
{(2n+1)!}Q_{-1/2}\!\left[\sigma_{2n+1}(R_1,R_2)\right]\!\right\}\!.
\hspace*{.3cm}&&
\end{eqnarray}
Here, we have introduced the notation
\begin{eqnarray*}
\Delta t=\frac{t_2-t_1}{4\bar{G}},\hspace{.5cm}
\sigma_n(R_1,R_2)=\frac{16\bar{G}^2n^2+R_1^2+R_2^2}{2R_1R_2}.
\end{eqnarray*}
Besides, $Q_{-1/2}(x)=\pi
F\left(\frac{3}{4},\frac{1}{4};1;\frac{1}{x^2}\right)/\sqrt{2x}$
[with $x>1$] is the associated Legendre function of the second
kind \cite{Gradshteyn:1994gr}. We recall that the function
$Q_{-1/2}(x)$ grows without bound as the argument approaches $x=1$
and falls off to zero as $\pi/\sqrt{2x}$ when
$x\rightarrow\infty$. When $R_1=R_2$ the singularity in Eq.
(\ref{s2-006}) comes just from the $n=0$ term in the first series
of the expansion, that is given by
$$\frac{i8\bar{G}}{\pi\sqrt{R_1R_2}}\,
Q_{-1/2}\left(\frac{R_1^2+R_2^2}{2R_1R_2}\right)
\sin\left(\frac{t_2-t_1}{4\bar{G}}\right).
$$
A series of plots of
$[\hat{\psi}(R_1,T_1=t_1),\hat{\psi}(R_2,T_2=t_2)]$ and $\langle
0|\, [\hat{\psi}_E(R_1,t_1),\hat{\psi}_E(R_2,t_2)] \,|0\rangle$
(both over $8\bar{G}$) is shown in Fig. \ref{fig4:compare} for
fixed values of $R_1$ and $t_2-t_1$ as a function of $R_2$, with
several choices for $\bar{G}$. We choose $t_2-t_1$ small enough to
guarantee the rapid convergence of the series in Eq.
(\ref{s2-006}) and leave a discussion of the behavior of the
integral (\ref{s2-005}) when $\bar{G}\rightarrow0$ for future
work. As we can see $\langle 0|\,
[\hat{\psi}_E(R_1,t_1),\hat{\psi}_E(R_2,t_2)]\,|0\rangle$ seems to
approach $[\hat{\psi}(R_1,T_1=t_1),\hat{\psi}(R_2,T_2=t_2)]$ at
least in a certain average sense when $\bar{G}$ is sufficiently
small (though not vanishing). It falls off to zero quite quickly
outside the light cone defined by the free commutator and the
auxiliary Minkowski metric, and displays an oscillatory behavior
within this light cone. The characteristic length of this
oscillation decreases with $\bar{G}$, as well as close to the
$R_1=R_2$ singularity. The approximation obtained by truncating
the series expansion (\ref{s2-006}), keeping a sufficiently large
number of terms, compares well with the results of numerically
computing expression (\ref{s2-005}), at least for low enough
values of $t_2-t_1$.
\bigskip

\noindent$\bullet$ Expectation value on one-particle states.
\bigskip

We consider now states of the form
$$
|\rho\rangle=\int_0^\infty \!\!\!\!dk \,f(k)A^\dagger (k)|0\rangle
$$
where the function $f=|f|e^{i\phi_f}$ satisfies $\int_0^\infty \!
dk|f(k)|^2=1$. We then have
\begin{eqnarray}
&&\langle \rho
|[\hat{\psi}_E(R_1,t_1),\hat{\psi}_E(R_2,t_2)]|\rho\rangle=
\nonumber \\ && -i8\bar{G}\!\int_0^\infty\!\!\!\!\!\!dk_1\!\!
\int_0^\infty\!\!\!\!\!\!dk_2 J_0(R_1k_1)\bigg\{2
J_0(R_2k_2)|f(k_1)f(k_2)|\nonumber\\ && \times\sin{\left[
2\bar{G}(t_1-t_2)E(k_1)E(k_2)\right]}
\cos{\left[\Omega_f(k_1,k_2)\right]} \nonumber\\ &&
-J_0(R_2k_1)|f(k_2)|^2\sin{\left[(t_2-t_1)E(k_1) e^{-4\bar{G}k_2}
\right]}\bigg\},\hspace*{.9cm} \label{s2-007}
\end{eqnarray}
where
\begin{eqnarray}
\Omega_f(k_1,k_2)&=&2\bar{G}(t_2-t_1)E(k_1)E(k_2)+t_1E(k_1)\nonumber\\
&-&t_2E(k_2)-\phi_f(k_1) +\phi_f(k_2).\hspace*{.4cm}\nonumber
\end{eqnarray}

A complete discussion of the meaning of the previous expression is
beyond the scope of this paper. Nevertheless, some features
already present in the vacuum expectation value are also present
here; in particular the $R_1=R_2$ singularity. This can be seen by
considering the last term in Eq. (\ref{s2-007}): the integral in
$k_2$ is
$$
\int_0^{\infty}\!\!\!dk_2|f(k_2)|^2\sin{\left[(t_2-t_1)E(k_1)
e^{-4\bar{G}k_2}\right]},
$$
which takes in general a non-vanishing constant value (depending
on $t_2-t_1$ and $\bar{G}$) as $k_1\rightarrow\infty$, thus
rendering the remaining integral in $k_1$ divergent. As the first
term in Eq. (\ref{s2-007}) leads to a convergent integral, we
conclude that the expectation value is singular when $R_1=R_2$.

It is not difficult to obtain as well an explicit expression for
the expectation value of the $E(H_0)$-commutator on the coherent
states of the field $\psi$. These diagonal matrix elements are
calculated in Appendix B. For our discussion in this work, let us
only comment that the result is actually divergent when $R_1=R_2$.
This supplies further support to the claim that the considered
singularity is indeed a generic feature of the system.


\section{\label{Conclusions}Conclusions and perspectives}

Linearly polarized cylindrical waves can be studied in great
detail both from the classical and quantum points of view. As we
have seen, there are two relevant Hamiltonians for the study of
the system. We have shown that the action and the metric of the
gauge-fixed model in linearized gravity reproduce the results
obtained by considering full cylindrical gravity and working to
the first perturbative order. We get in this way a free
Hamiltonian. The Hamiltonian governing the dynamics of the full
system, on the other hand, is different from the free one, but
turns out to be a function of it and presents certain features
with deep physical consequences, like e.g. the existence of an
upper bound.

We have studied the similarities and differences of these two
admissible kinds of evolution; in particular, we have discussed
how the emergence of an upper bound for the energy affects the
causal structure of the model and the spreading of the light
cones. The field commutator for the free Hamiltonian is a c-number
and shows the typical light cone structure found in standard
perturbative quantum field theories. The commutator for the
physical Hamiltonian, as it usually happens for interacting
theories, is no longer a c-number, so one has to consider its
matrix elements. By concentrating on the vacuum expectation value
we have been able to see several interesting phenomena: a
spreading of the light cone as a function of the gravitational
constant, the disappearance of the singularity present in the free
case due to the smearing of the symmetry axis and the appearance
of a new type of singularity associated with the fact that the
energy is bounded from above. This new singularity is also present
in the other expectation values discussed in the paper, namely for
one-particle states and coherent states, and appears to be a
generic feature of the model.

There are several open questions that we plan to address in future
work. In particular, it would be desirable to reach a better
understanding of the behavior of the field commutator in the limit
in which the length scale provided by $\bar{G}$ goes to zero. The
expectation values of the $E(H_0)$-commutator discussed here
resemble those derived from the free Hamiltonian $H_0$ at least in
a certain average sense. However it is not obvious how precisely
and up to what extent they actually relate to each other. This is
partly so because of the different singularity structure found in
both cases. Further research on this subject will concentrate on
the properties of the model in the semiclassical limit
$\bar{G}\rightarrow 0$. We will also pay detailed attention to
matrix elements of the field commutator other than the vacuum
expectation value, with the aim at discussing how the smearing of
the light cones depends on the energy.

\appendix

\section{\label{A1}Useful Commutators}

In this appendix, we compute the commutators of the creation and
annihilation operators $\hat{A}_E(k,t)$ and
$\hat{A}^{\dagger}_E(k,t)$, obtained from the corresponding
operators $\hat{A}(k)$ and $\hat{A}^{\dagger}(k)$ via the unitary
evolution generated by $E(\hat{H}_0)$, where
$\hat{H}_0=\int^{\infty}_0dkk\hat{A}^{\dagger}(k)\hat{A}(k)$.
Employing relations (\ref{crean2}) and the basic commutators
(\ref{comma}), it is possible to show that
\begin{widetext}
\begin{eqnarray*}
\bullet \, \big[\hat{A}_E(k_1,t_1),\hat{A}_E(k_2,t_2)\big]&=&
\hat{A}(k_1)\hat{A}(k_2)
\exp\left[it_1E(\hat{H}_0-k_1-k_2)+i(t_2-t_1)E(\hat{H}_0-k_2)
-it_2E(\hat{H}_0)\right]
\\&-&\hat{A}(k_1)\hat{A}(k_2)\exp\left[it_2E(\hat{H}_0-k_1-k_2)
+i(t_1-t_2)E(\hat{H}_0-k_1)-it_1E(\hat{H}_0)\right]\,,
\\
\bullet \,
\big[\hat{A}^\dagger_E(k_1,t_1),\hat{A}^\dagger_E(k_2,t_2)\big]&=&
\hat{A}^\dagger(k_1)\hat{A}^\dagger(k_2)
\exp\left[it_1E(\hat{H}_0+k_1+k_2)+i(t_2-t_1)E(\hat{H}_0+k_2)
-it_2E(\hat{H}_0)\right]
\\& -&\hat{A}^\dagger(k_1)\hat{A}^\dagger(k_2)
\exp\left[it_2E(\hat{H}_0+k_1+k_2)+i(t_1-t_2)E(\hat{H}_0+k_1)
-it_1E(\hat{H}_0)\right]\,,
\\
\bullet \,
\big[\hat{A}_E(k_1,t_1),\hat{A}^\dagger_E(k_2,t_2)\big]&=&
\hat{A}(k_1)\hat{A}^\dagger(k_2)
\exp\left[it_1E(\hat{H}_0+k_2-k_1)+i(t_2-t_1)E(\hat{H}_0+k_2)
-it_2E(\hat{H}_0)\right]
\\&-&\hat{A}^\dagger(k_2)\hat{A}(k_1)\exp\left[it_2
E(\hat{H}_0+k_2-k_1)
+i(t_1-t_2)E(\hat{H}_0-k_1)-it_1E(\hat{H}_0)\right]\,,
\\
\bullet \, \big[\hat{A}^\dagger_E(k_1,t_1),A_E(k_2,t_2)\big]&=&
\hat{A}^\dagger(k_1)\hat{A}(k_2)
\exp\left[it_1E(\hat{H}_0+k_1-k_2)+i(t_2-t_1)E(\hat{H}_0-k_2)
-it_2E(\hat{H}_0)\right]\\&-&\hat{A}(k_2)\hat{A}^\dagger(k_1)
\exp\left[it_2E(\hat{H}_0+k_1-k_2)
+i(t_1-t_2)E(\hat{H}_0+k_1)-it_1E(\hat{H}_0)\right].
\end{eqnarray*}
\end{widetext}

\section{\label{A2}Expectation values on coherent
states}

We consider coherent states of the field $\psi$, given by
\begin{eqnarray*}
|\Psi_C\rangle&=&K_C\exp{\left(\int_0^\infty\!\!\!\frac{dk}
{\sqrt{8\bar{G}}}\,C(k)A^\dagger(k)\right)}|0\rangle \\
&=&K_C\sum_{n=0}^\infty\frac{1}{n!}\left(\int_0^\infty\!\!\!
\frac{dk}{\sqrt{8\bar{G}}}\,C(k)A^\dagger (k)\right)^n|0\rangle,
\end{eqnarray*}
where $C(k)$ is a square integrable function and $K_C$ is a
normalization constant satisfying
\begin{eqnarray*}
|K_C|^2=\exp\left(-\int_0^\infty\!\frac{dk}
{8\bar{G}}\,|C(k)|^2\right).
\end{eqnarray*}
The expectation value of the $E(H_0)$-commutator is
\begin{eqnarray*}
\langle
\Psi_C|\big[\hat{\psi}_E(R_1,t_1),\hat{\psi}_E(R_2,t_2)\big]
|\Psi_C\rangle=\hspace{2.4cm}\\
\int_0^\infty\!\frac{dk_1}{2}\!\!\int_0^\infty \!\!\!\!\!\!dk_2
J_0(R_1k_1)J_0(R_2k_2)\sum_{s=0}^\infty\frac{I_s(C)}
{s!}G_s(k_1,k_2),\end{eqnarray*} where
\begin{eqnarray*}
I_s(C)=\exp\left\{\int_0^\infty\!\frac{dk}{8\bar{G}}\,|C(k)|^2
\left(e^{-4\bar{G}s k}\!-\!1\right)\right\},\hspace*{1.2cm}\\
G_s(k_1,k_2)=8\bar{G}\delta(k_1,k_2) \bigg\{[b(k_1,-k_2)]^s
\!\!-\![b(k_2,-k_1)]^s\!\bigg\}\\+
\overline{C}(k_2)C(k_1)\bigg\{[b(k_1,-k_2)]^s\!-
\![b(-k_2,k_1)]^s\bigg\}e^{-4\bar{G}s k_1}\hspace*{.2cm}\\
+\overline{C}(k_1)C(k_2)\bigg\{[b(-k_1,k_2)]^s\!
-\![b(k_2,-k_1)]^s\bigg\}e^{-4\bar{G}s k_2}\hspace*{.2cm} \\
+C(k_1)C(k_2)
\bigg\{[b(k_1,k_2)]^s\!-\![b(k_2,k_1)]^s\bigg\}e^{-4\bar{G}s
(k_1+k_2)}
\\ + \overline{C}(k_1)\overline{C}(k_2)
\bigg\{[b(-k_1,-k_2)]^s\!-\![b(-k_2,-k_1)]^s\bigg\},
\hspace*{.7cm}
\end{eqnarray*}
and we have employed the notation
\begin{eqnarray*}
b(k_n,k_m)&=&\!\frac{-i}{4\bar{G}}\left[ t_ne^{4\bar{G}
k_m}(e^{4\bar{G}k_n}\!-\!1)+t_m(e^{4\bar{G} k_m}\!-\!1)\right].
\end{eqnarray*}
Note that, when $R_1=R_2$, the delta in the expression of
$G_s(k_1,k_2)$ leads to the divergent integral
\begin{eqnarray*}
\int_0^\infty\!\!\!\!\!dk\,
J_0^2(R_1k)\!\sum_{s=0}^\infty\frac{8\bar{G}I_{2s+1}}
{(2s+1)!}\left[\frac{i(t_2-t_1)
(1-e^{-4\bar{G}k})}{4\bar{G}}\right]^{\!2s+1}\!\!\!\!\!\!\!\!.
\end{eqnarray*}

\begin{acknowledgments} The authors wish to thank A.
Ashtekar, G. Date, and L. Garay for interesting discussions. They
are especially grateful to M. Varadarajan for suggesting the
subject and sharing enlightening conversations and insight.
E.J.S.V. is supported by a Spanish Ministry of Education and
Culture fellowship co-financed by the European Social Fund. This
work was supported by the Spanish MCYT under the research projects
BFM2001-0213 and BFM2002-04031-C02-02.
\end{acknowledgments}

\end{document}